# Landau Level Anticrossing Manifestations in the Phase Diagram Topology of a Two Subband System

X. C. Zhang, I. Martin[*], and H. W. Jiang

*Department of Physics and Astronomy, University of California at Los Angeles,
405 Hilgard Avenue, Los Angeles, CA 90095, USA*
[*]*Theoretical Division, Los Alamos National Laboratory, Los Alamos, New Mexico 87545, USA*

## Abstract

In a two-subband GaAs/AlGaAs two-dimensional electron system, the phase diagram of longitudinal resistivity $\rho_{xx}$ in density and magnetic field plane exhibits an intriguing structure centered at filling factor $\nu = 4$ which is strikingly different from the ring-like structures at lower magnetic fields. Thermal activation measurements reveal an anticrossing gap on each boundary of the structure where intersubband Landau Levels with parallel or antiparallel spin are brought into degeneracy. While the physics of the anticrossing can be ascribed to the pseudospin quantum Hall ferromagnetism, as reported earlier by Muraki et al, the mapping and modeling of the phase diagram topology allow us to establish a more complete picture of the consequences of real spin / pseudo-spin interactions for the two subband system.

A two-subband electron system, which can be either a wide single quantum well or a coupled double quantum well, becomes particularly appealing for studying correlation effects when the two sets of Landau levels (LLs) are brought into degeneracy by varying magnetic field or carrier concentration [1]. The correlation effects can be intriguing near these degeneracy points, since in addition to the interaction of the real electron spins, there is also a strong correlation of the two subband charge states, referred in the literature as pseudospins [2-6]. For example, in an earlier study it was found that the experimentally determined phase diagram of a single well two-subband system, exhibits "ring-like structures" at even integer filling factors [7] near degeneracy points. Inside the rings, due to the exchange interactions of real spins, the electron spins of different subbands are aligned by the external magnetic field over an extended range of magnetic field and density [8,9]. On the other hand, an early work in a wider single well two subband system, energy gaps were observed near the degeneracy points at filling factors of both $\nu=3$ and 4 [10], which were interpreted as a result of the easy-plane or easy-axis quantum Hall ferromagnetism of the pseudospin charge states.

Here, we extend the study of the phase diagram topology to higher magnetic fields up to 10 T, with the main objective to explore the interplay between the real spin and the pseudospin correlations. We found that the "ring-like" structures at low magnetic fields, centered around $\nu = 6, 8, 10$, evolve into a "square-like" structure at filling factor $\nu = 4$ at higher magnetic fields which shows clear evidence of level anticrossing at its four boundaries. The evidence was established by both thermal activation gap measurement in the proximity of the expected LL degeneracy points, and also by reproducing key features in the phase diagram topology using simulation of the density of states (DOS) of the anticrossed LLs. We believe the anti-crossings exhibited in our phase diagram are consistent with the early reported quantum Hall ferromagnetism of pseudospins [10].

The sample used in this work is a symmetrical modulation-doped single quantum well with an equally 240 Å thick well and spacer. The sample details can be found in Ref. 7. At zero gate voltage the total electron density is $8.1 \times 10^{11}$ cm$^{-2}$, the density in the ground and first excited subband are $5.4 \times 10^{11}$ cm$^{-2}$, $2.7 \times 10^{11}$ cm$^{-2}$, respectively. The mobility is about $4\times10^5$ cm$^2$/V.s. The spin splitting is well resolved above 1.8 Tesla for both subbands. The band edge profile and subband wave function envelopes are depicted in Fig. 1b. A 100 μm wide Hall bar with 270 μm between voltage probes was patterned by standard lithography techniques. A NiCr top gate was used to vary the electron density. Measurements were made in a dilution refrigerator with a base temperature of 60 mK and in a perpendicular magnetic filed normal to sample plane using ac lock-in techniques with a bias current of 10 nA.

The phase diagram, i.e., gray scale plot of longitudinal resistivity $\rho_{xx}$ as a function of magnetic field $B$ and gate voltage $V_g$, is shown in Fig.1a. The most pronounced feature in this diagram is the last "square-like structure" enclosed in the dashed line box, strikingly different from the reported ring-like structures at low magnetic fields [7]. The most noticeable feature of the square-like structure is the *disappearance* of the extended states on its four boundaries, marked by "A", "B", "C", and "D" in Fig. 1a. These are the places where spin split LLs of different subbands cross each other. The evolution from a ring to square can be partly seen from the gradual disappearance of the extended states at the lower branch of the ring indicated by the three arrows. The labeled filling factors are read from the quantized values in units of $e^2/h$ in transverse conductivity $1/\rho_{xy}$ simultaneously measured with $\rho_{xx}$.

Since there seems to exist a gap manifested by the disappearance of the extended states on each boundary of the square structure, it is natural to obtain a measurement of the energy scale of the associated gap size there. In Fig. 2b we present the Arrehenius plots with $log(\rho_{xx})$ as a function of inverse temperature $1/T$ for four data points "A, B, C, D" marked by dots in Fig. 2a, which are center of each boundary of the square structure. The Arrehenius plots all show clear temperature activation behavior $\rho_{xx} \propto \exp(-E_a/2T)$, and obtained activation energy $E_a$ are 2.70, 2.88, 3.34, 2.86 K for "A", "B", "C" and "D", respectively. The evolution of the gap at $v = 3$ and 4 along the dashed line in Fig. 2a are shown in Fig. 2c. Both curves indicate an *anticrossing gap* signified by a minimum at "A" and "B", respectively. Apparently, the gap at $v = 4$ undergoes a much more rapid transition as a function of magnetic field.

We believe the anticrossings, or energy gaps, observed here have the same origin as those reported earlier by Muraki et al in a similar two subband system [10]. The authors suggested that the minima in the activation energy are results of pseudospin quantum Hall ferromagnetism, which lifts the degeneracy near the Landau level crossing points. It has been recognized in recent years that when two Landau levels with different symmetry of their wave functions, labeled as pseudospin up and down states, simultaneously approach the chemical potential at integer filling factors [2], ferromagnetic state of the pseudospins becomes energetically favorable. The resultant pseudospin anisotropy energy depends on the details of both orbital and spin states of LLs involved, accordingly the system can exhibit easy axis or easy plane anisotropy. The activation gap in the longitudinal resistivity corresponds to the energy required to create an unbound particle-hole excitation in such an insulating ferromagnetic state. Following the analysis of Ref. 9, the activation energy $E_a$ is replotted versus the effective Zeeman energy $\Delta E_z$ in Fig. 3 for $\nu$ = 4 and 3, respectively, both normalized by characteristic Coulomb energy $e^2/4\pi\varepsilon \cdot l_B$, where $l_B = (\hbar/eB)^{1/2}$ is magnetic length. The single particle energy difference $\Delta E_z$ acts as *effective* Zeeman energy,

$$\Delta E_z = (|\Delta N|\frac{\hbar e}{m^*} + \Delta\sigma|g|\mu_B) \times (B - B_C), \qquad (1)$$

$\Delta N$ or $\Delta\sigma$ is LL or spin index change at crossing field $B_c$. $\Delta N$ = 1(1) and $\Delta\sigma$ = ±1 (0) for LL crossing at $\nu$ = 4 (3). $g$ = -0.44 is the bare g-factor. The gaps for $\nu$ = 3 in Fig. 3b show a smooth transition whose asymptotes approach the single particle $\Delta E_z$, indicated by the straight lines. This is consistent with a continuously evolving ground state which is a coherent superposition of two-subband pseudospins, or a easy-plane magnetization anisotropy [2,10]. The magnitudes of the gaps are consistent with the theoretical quasiparticle energy gap calculated by Jungwirth et.al for idealized easy-plane anisotropy [3]. On the contrary, normalized $E_a$ at $\nu$ = 4 in Fig. 3a shows a slope of 4 times greater than the single particle Zeeman gap denoted by straight lines. This unusual behavior is likely to be caused by the easy-axis ferromagnetism. The high slope may be caused by the disorder-broadened gap hysteresis predicted by Jungwirth et al [3] for the easy axis quantum Hall ferromagnets. While theory predicts a first order transition between fully occupied subband levels in this case, in a real system disorder will lead to finite-size domain formation, possibly softening the bulk transition to the second order, which is consistent with the smooth evolution of the activation gap in the vicinity of the $v = 4$ anticrossing. We note here that the slope enhancement that we observe is a factor of 10 smaller than in Muraki et al which is likely caused by the difference in the system parameters, including the well width, carrier densities and disorder.

It is instructive and interesting to check if we can understand the intriguing topological features of the phase diagram within the LL anticrossing picture. To that end we developed a simple analytical model based on DOS calculation of anticrossed spin split LLs with different subband indices. It turns out that by properly choosing the anticrossing gap, the main features of the phase diagram can be reproduced quite nicely. For a noninteracting two subband system in a perpendicular magnetic field $B$, the energy spectrum is,

$$E_{1,\pm} = (n+\frac{1}{2})\hbar\omega_B \pm g^*\mu_B B \qquad (2a)$$

$$E_{2,\pm} = (n+\frac{1}{2})\hbar\omega_B \pm g^*\mu_B B + \Delta E_{12} \qquad (2b)$$

$E_{1,\pm}$ and $E_{2,\pm}$ are energy of ground and first excited state with different spin states, $\hbar\omega_B$ is cyclotron energy, $g^*$ is electron effective g-factor, $\mu_B$ is Bohr magneton, $\Delta E_{12}$ is the energy offset between the two subbands. For level anticrossing, here we choose an energy dispersion of a generic square root form irrespective of the spin states,

$$E_A = \frac{(E_1+E_2)}{2} + \sqrt{\frac{1}{4}(E_1-E_2)^2 + \Delta^2} \qquad (3a)$$

$$E_B = \frac{(E_1+E_2)}{2} - \sqrt{\frac{1}{4}(E_1-E_2)^2 + \Delta^2} \qquad (3b)$$

$E_A$, $E_B$ are antibonding and bonding states after level repulsion caused by anticrossing, $E_1$ and $E_2$ are single particle energy levels expressed by Eq. (2a) and (2b). $\Delta$ is half of the anticrossing gap between two spin split LLs of different subbands. In order to directly compare with experiment, we need to calculate DOS of the anticrossed LLs. For simplicity, we assume the resistivity, that is proportional to the DOS, takes a Lorentzian form. At a fixed magnetic field, for a certain electronic density $n$, the Fermi energy $E_F$ can be determined by integrating the Lorentzian DOS over the occupied states ($E \leq E_F$) and making it equal to $n$. The DOS (i.e, the resistivity) can be deduced from this $E_F$. Repeating the procedure for different $n$ and $B$, the whole topological phase diagram can be constructed as shown in Fig. 4 [11]. The top panel of Fig. 4a shows the result of zero anticrossing, i.e., the LLs just cross each other like in normal case. The ring structures closely match the experimental diagram in Fig. 1a at low magnetic fields, which supports the validity of our model. As expected, because of the overlap of DOS, at each LL crossing point such as "A", "B", "C", and "D" shown in the inset, the resistance value has a maximum marked by the arrows in the dashed line box. To account for the exchange interaction of the real spins, we have used an enhanced effective g-factor $g^* = -2.2$ for each subband, similar to that used by Ellenberger et. al recently [12]. It is apparent that this normal LL crossing picture fails completely to explain the conductance disappearance at the LL crossing points marked by "A, B, C, D" in Fig. 2a. However, by using the experimental gap values ($\approx 0.12$ meV at 7.725 T) for $\Delta$ and assume it is a linear function of $B^{1/2}$ in Eq. 3, the square-like structure readily comes out in high $B$ as shown in Fig. 4b. This exercise clearly indicates that the intersubband LL anticrossing effect is responsible for the "square-like" structure in topological phase diagram.

Having mapped and modeled the phase diagram topology, we may now gain some new insight for a better global understanding of the two-subband system. For example, it can be anticipated that along ν = 5 anticrossing should also occur which shares the same physics as ν = 3 case. Second, the topological phase diagram in Fig. 1a displays interesting interplay between the real spin and the pseudospin correlations as $B$ increases. Along even filling factor line sectioning through the ring (or square) structure, as shown in Fig. 2a, it represents a paramagnetic to ferromagnetic phase transition induced by preferential alignment of *real* electron spins to save exchange energy in an external magnetic field [7-9]. At high $B$, *in addition,* the pseudospin correlation resultant easy-axis or easy-plane quantum Hall ferromagnetism at even or odd filling factors, respectively, opens up a gap at each boundary of the "ring" structure making it appear like a "square" at ν = 4. Interestingly enough, the gap at odd filling factors appear to open up gradually with $B$ evidenced by the gradual disappearance of the resistance at ν = 7, 5 and 3 shown by the three arrows in Fig. 1a. In contrast, the gap occurred on even filling factor boundary seems to open up suddenly at $\nu = 4$. The *B*-dependence of the four anti-crossing gaps around the square (or ring) structure at $\nu = 4, 6, 8$ is shown in Fig. 2d. We speculate that the very different *B*-dependence of the gaps at even and odd filling factors reflects the difference in the disorder effects in the easy-plane and easy- axis anisotropy cases. For instance, potential out-of-plane disorder can cause fluctuations in the position of the subband crossing point. While in the easy plane case, the electronic state can gradually adjust to such fluctuations via a smooth distortion, in the easy axis case, strong enough disorder fluctuations can make the structure with domains of fully occupied subband levels unsustainable, thus eliminating the single-particle gap. This speculation, however, can only be validated by a further quantitative theoretical analysis.

In conclusion, in a two-subband GaAs/AlGaAs two-dimensional electron system, topological diagram of resistivity in density and magnetic field plane evolves from ring-like structures into a square-like structure at higher magnetic fields. On each boundary of the square structure where spin split intersubband LLs are degenerate, an anticrossing gap was revealed by the thermally activated behavior. The simulation of the phase diagram topology based on DOS calculation of anticrossed intersubband LLs shows good qualitative agreement with experiment. We attribute the origin of the gaps to pseudospin-ferromagnetism as reported before.

The authors would like to thank K. Yang for helpful discussions, and B. Alavi for technical assistance. This work is supported by NSF under grant DMR-0404445.

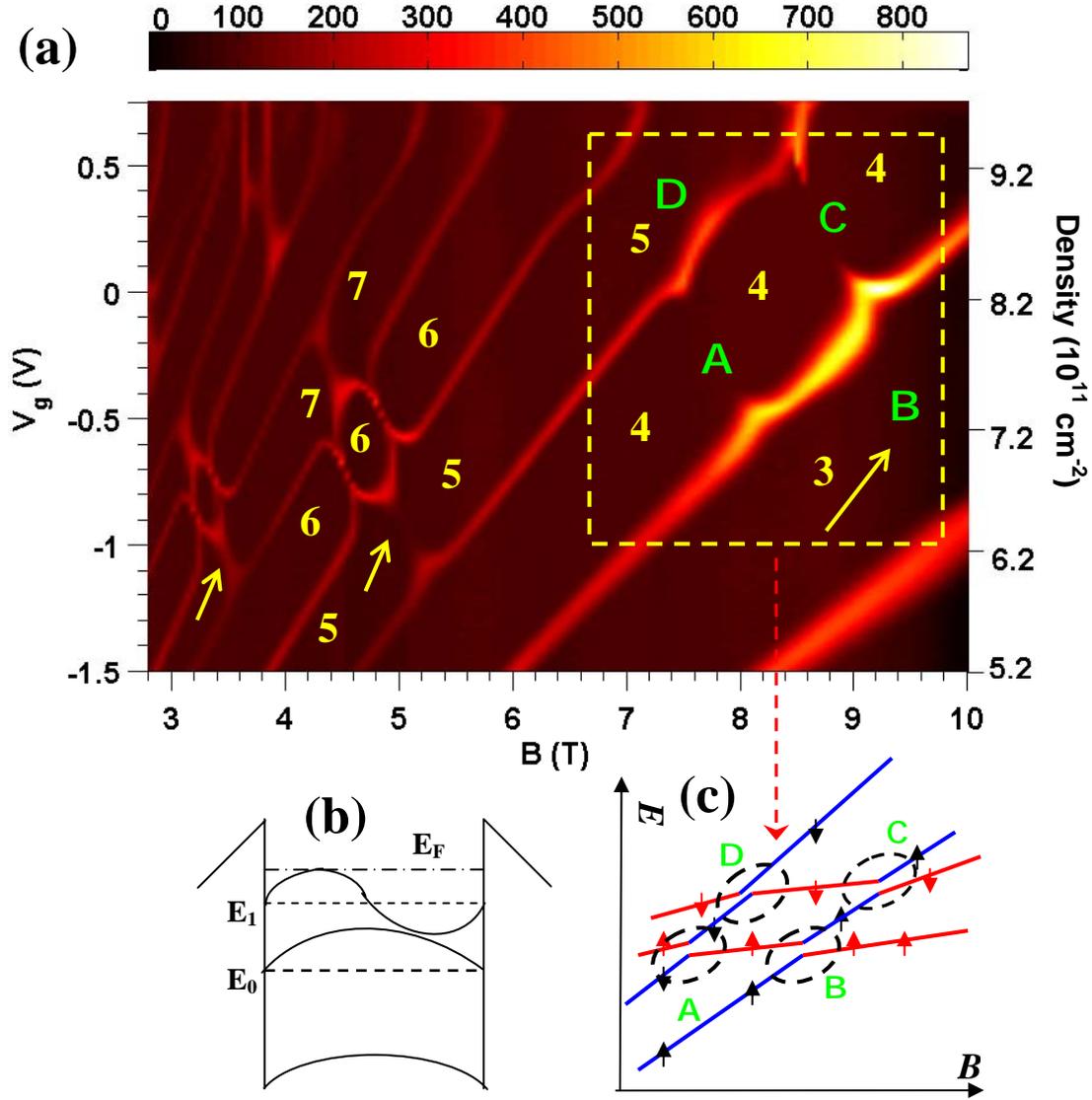

Fig. 1 (a), Gray-scale plot of the longitudinal resistivity $\rho_{xx}$ in the gate-voltage $V_g$ and magnetic field $B$ plane measured at 70 mK. The filling factors from $\rho_{xy}$ measurement are labeled. A square structure at $\nu = 4$ (inside the dashed line box) is observed at high magnetic fields, in contrast to the ring structures at low magnetic fields. Fig. (b) schematically shows band edge profile, energy quantization, and wave function envelope of our two-subband system. Fig. (c) depicts anticrossing occurred between spin split intersubband LLs on four corresponding places in Fig. (a) marked by "A", "B", "C", and "D".

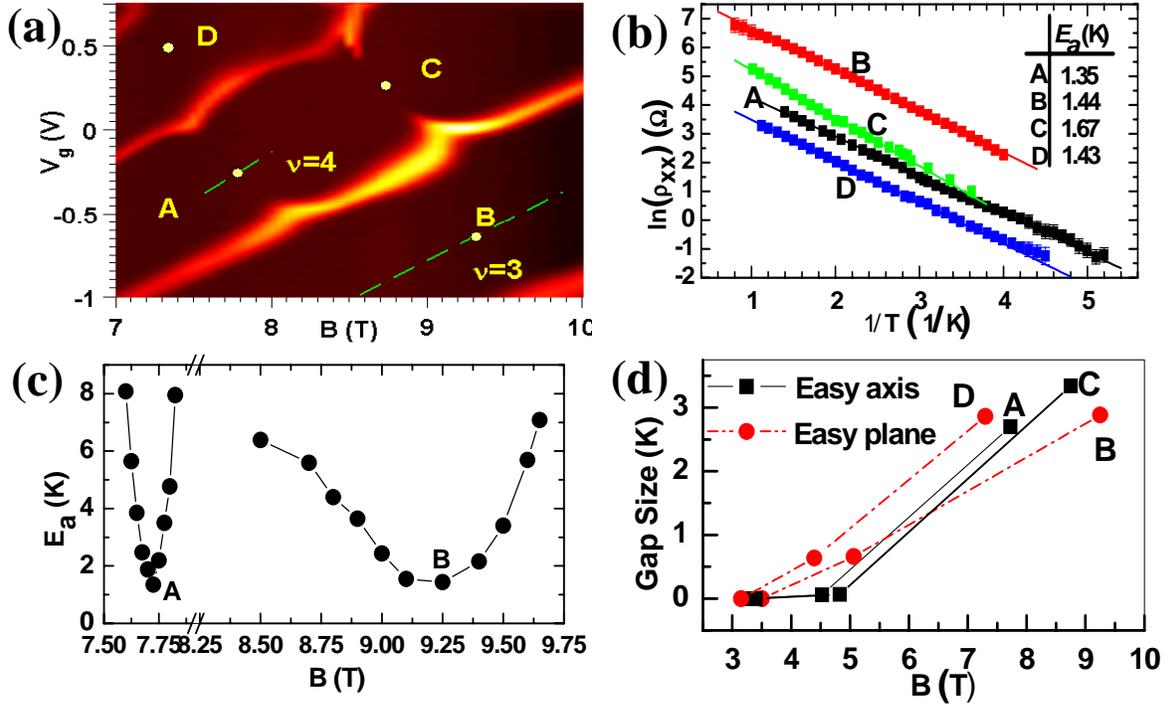

Fig. 2 (a), An enlarged view of the portion enclosed by box in Fig. 1a. Yellow dots A($V_g$=-0.207 V/$B$=7.725 T), B(-0.62 V/9.25 T), C(0.241 V/8.75 T), and D(0.505 V/7.3 T) are roughly middle point of each boundary of the ring where LLs are crossed. Fig. (b), Arrehenius plots of log($\rho_{xx}$) versus 1/$T$ for points A, B, C, and D. The straight lines are linear regression of experimental data. The activation energies are displayed in the inset table. Fig. (c), activation energy measured along the two dashed line centered at position "A" and "B" in (a) shows clearly anticrossing gap. Fig. (d), Anticrossing gap measured at positions "A, B, C, D" for square (or ring) structure centered at ν=4, 6, 8 as a function of $B$.

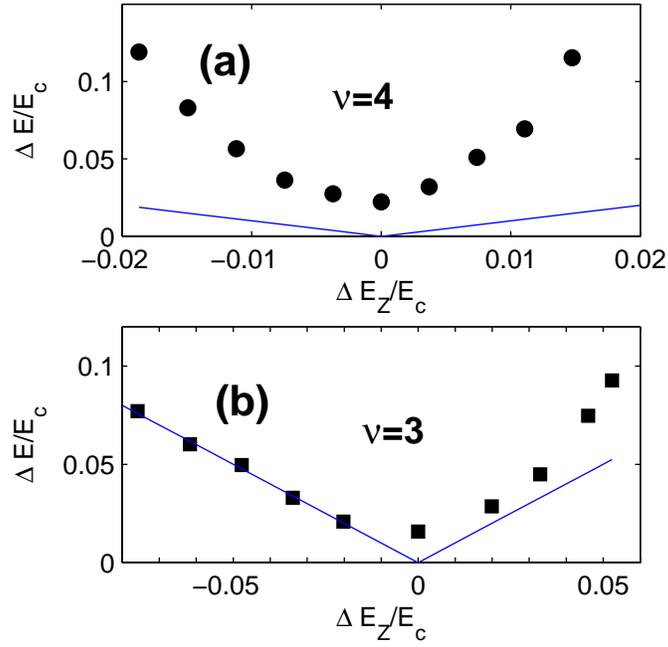

Fig. 3, Normalized experimental anticrossing gap $\Delta E$ (filled circles or squares) as a function of normalized single particle gap $\Delta E_z$, both in units of $e^2/4\pi\varepsilon l_B$, where $l_B = (\frac{\hbar}{eB})^{1/2}$ is the magnetic length, at filling factor $v = 4$ (a) and $v = 3$ (b). The lines in each figure have a slope of 1 and denote the effective single particle gap along the two dashed line shown in Fig. 2a.
11

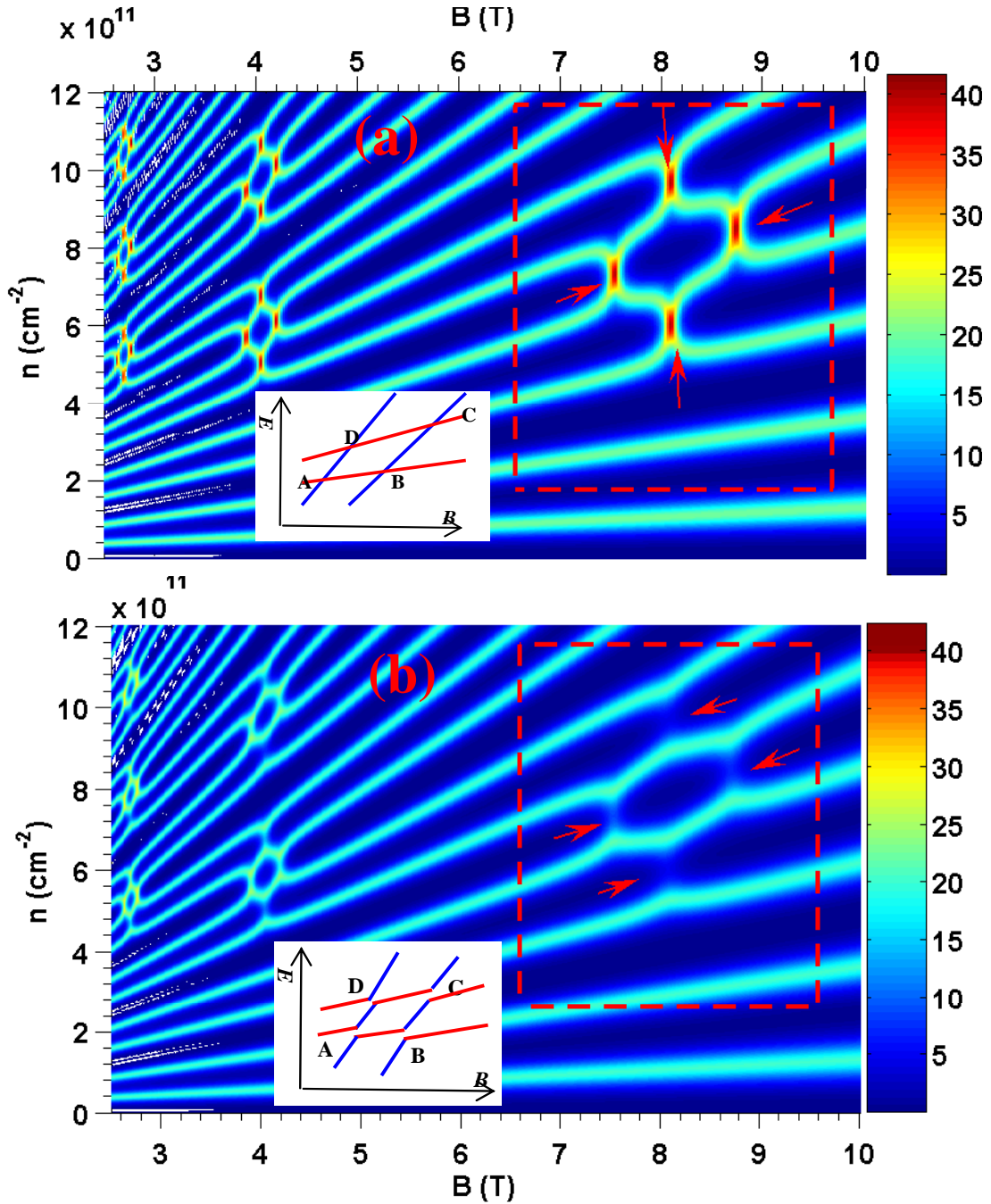

Fig. 4, Calculated resistivity (in arbitrary unit) topological diagram assuming zero anticrossing (a), and finite intersubband LL anticrossing gap Δ=0.12 meV at $B$=7.725 T (b). The contrast between them is emphasized in the dashed box. Instead of a resistivity enhancement for zero anticrossing case, four clear gaps appear at the four boundaries of the last ring structure for finite anticrossing (indicated by arrows in boxes). The inset in each graph shows corresponding LL crossing (a) or anticrossing (b) situation in $E$-$B$ plane.